\documentclass[12pt, preprint]{aastex}
\shorttitle{Diffuse UV Radiation from the Sandage Nebulosity}
\shortauthors{Sujatha et al.}

\newcommand{\phunit}{photons cm$^{-2}$ sr$^{-1}$ s$^{-1}$ \AA$^{-1}$}

\newcommand{\fuse} {{\it FUSE}}
\newcommand{\iras} {{\it IRAS}}
\newcommand{\galex} {{\it GALEX}}
\newcommand{\voyager} {{\it Voyager}}
\newcommand{\htwo} {H$_{2}$}

\begin{document}
\title{GALEX Observations of Diffuse UV Radiation at High Spatial Resolution from the Sandage Nebulosity}
\author{N. V. Sujatha, Jayant Murthy, Abhay Karnataki}
\email{sujatha@iiap.res.in, murthy@iiap.res.in, abhay@iiap.res.in}
\affil{Indian Institute of Astrophysics, Koramangala, Bangalore - 560 034, India}
\and
\author{Richard Conn Henry, Luciana Bianchi}
\email{henry@jhu.edu, bianchi@pha.jhu.edu}
\affil{Dept. of Physics and Astronomy}
\affil{The Johns Hopkins University, Baltimore, MD 21218}

\begin{abstract}

We have observed a region of nebulosity first identified as starlight scattered by interstellar dust by Sandage (1976) using the GALEX ultraviolet imaging telescope. Apart from airglow and zodiacal emission, we have found a diffuse UV background of between 500 and 800 \phunit\ in both the \galex\ FUV (1350 -- 1750 \AA) and NUV (1750 -- 2850 \AA) bands. Of this emission, up to 250 \phunit\ is due to \htwo\ fluorescent emission in the FUV band. The remainder is consistent with scattering from interstellar dust with forward scattering grains of albedo about 0.4. These are the highest spatial resolution observations of the diffuse UV background to date and show an intrinsic scatter beyond that expected from instrumental noise alone. Further modeling is required to understand the nature of this scatter and its implications for the ISM.

\end{abstract}


\section{INTRODUCTION}

Ever since the first observations of diffuse ultraviolet radiation by \citet{Hay69} and \citet{Lillie}, there has been an effort to understand its distribution and its origin. Unfortunately, because of the difficulty of the observations and the faintness of the background, many of the early observations were conspicuous more by their disagreements than by the light they shed on the topic. The state of the observations and theories before 1990 have been reviewed by \citet{Bowyer91} and \citet{RCH91}.

There has been significant progress in more recent years, particularly in the far ultraviolet ($\lambda < $ 1200 \AA) where \citet{JM99} and \citet{JM04} have used spectroscopic data from the \voyager\ and \fuse\ ({\it Far Ultraviolet Spectroscopic Explorer}) spacecraft, respectively, to trace the radiation field over many different locations in the sky. There have also been a number of observations at longer wavelengths, most recently by the SPEAR instrument \citep[][and references therein]{Ryu}, but no systematic study of the UV background. The {\it Galaxy Evolution Explorer} (\galex) offers us the opportunity to extend coverage of the diffuse background to a significant fraction of the sky with a sensitivity of better than 100 \phunit. In this work, we will report on one such observation: that of the nebulosity observed near M82 by \citet{Sandage76}, as a template for our further work with a much larger data set.

This cloud is at a high Galactic latitude (38\degr) with few nearby stars and was identified by \citet{Sandage76} as a canonical high latitude dusty cloud illuminated by the Galactic interstellar radiation field (ISRF). Our expectation was that we would be able to differentiate between starlight scattered from the Galactic cloud and extragalactic light which would be shadowed by the foreground cloud. The observations presented here are the first to probe the diffuse UV background at a spatial resolution comparable to other surveys of dust emission, notably the IR.

\section{OBSERVATIONS}

A full description of the \galex\ instrument and mission is given by \citet{Martin2005} and \citet{Morissey2007} with those parameters relevant to our observation listed in Table \ref{obslog}. Although we had originally intended for an exposure time of 15,000 seconds in each of the far ultraviolet (FUV) and near ultraviolet (NUV) bands, power problems in the FUV high voltage power supply and subsequent re-observations of the region gave us about 15,000 seconds exposure time in  the FUV and 35,000 seconds in the NUV. 

\begin{figure}
\epsscale{0.7}
\plotone{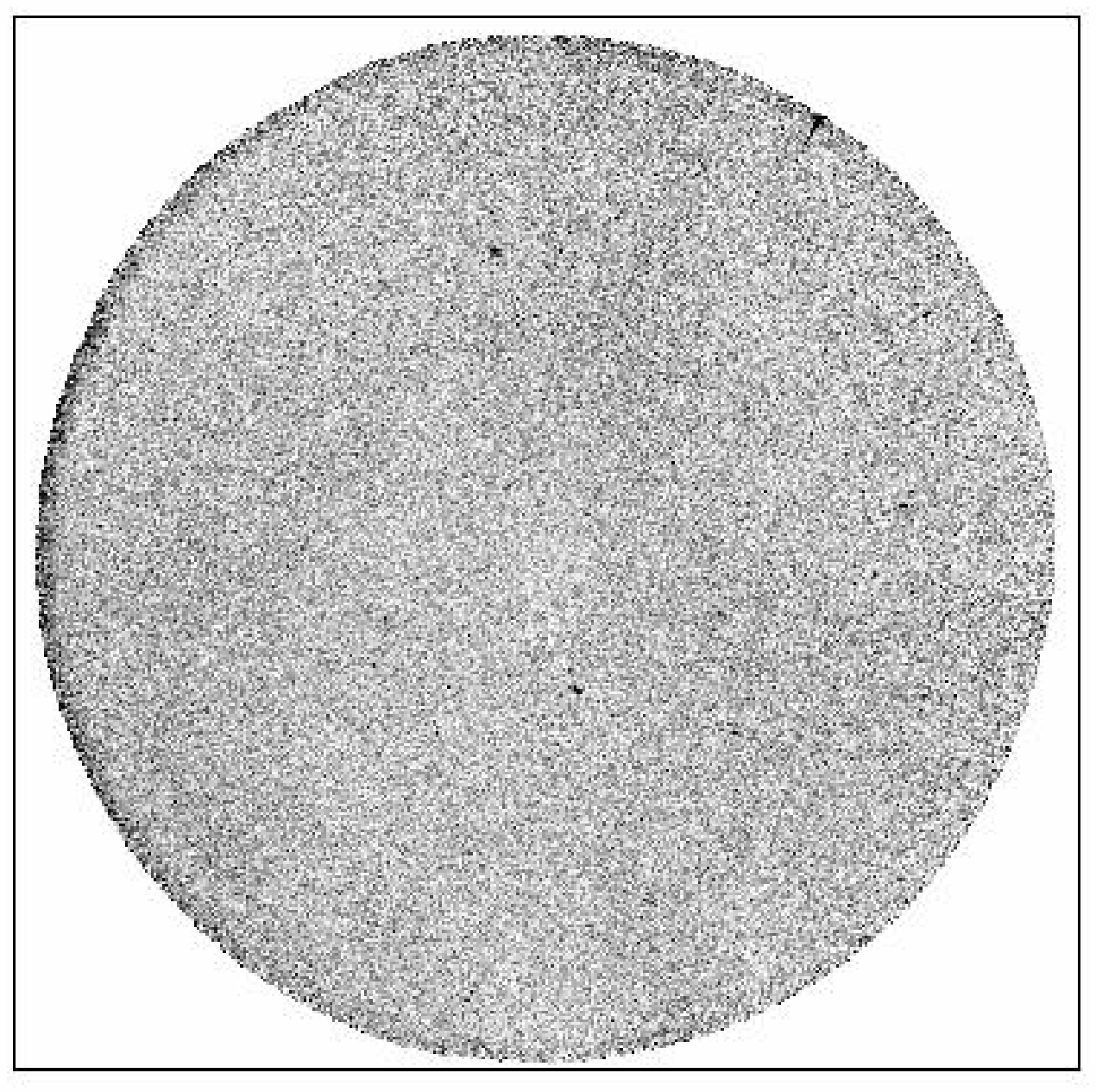}
\plotone{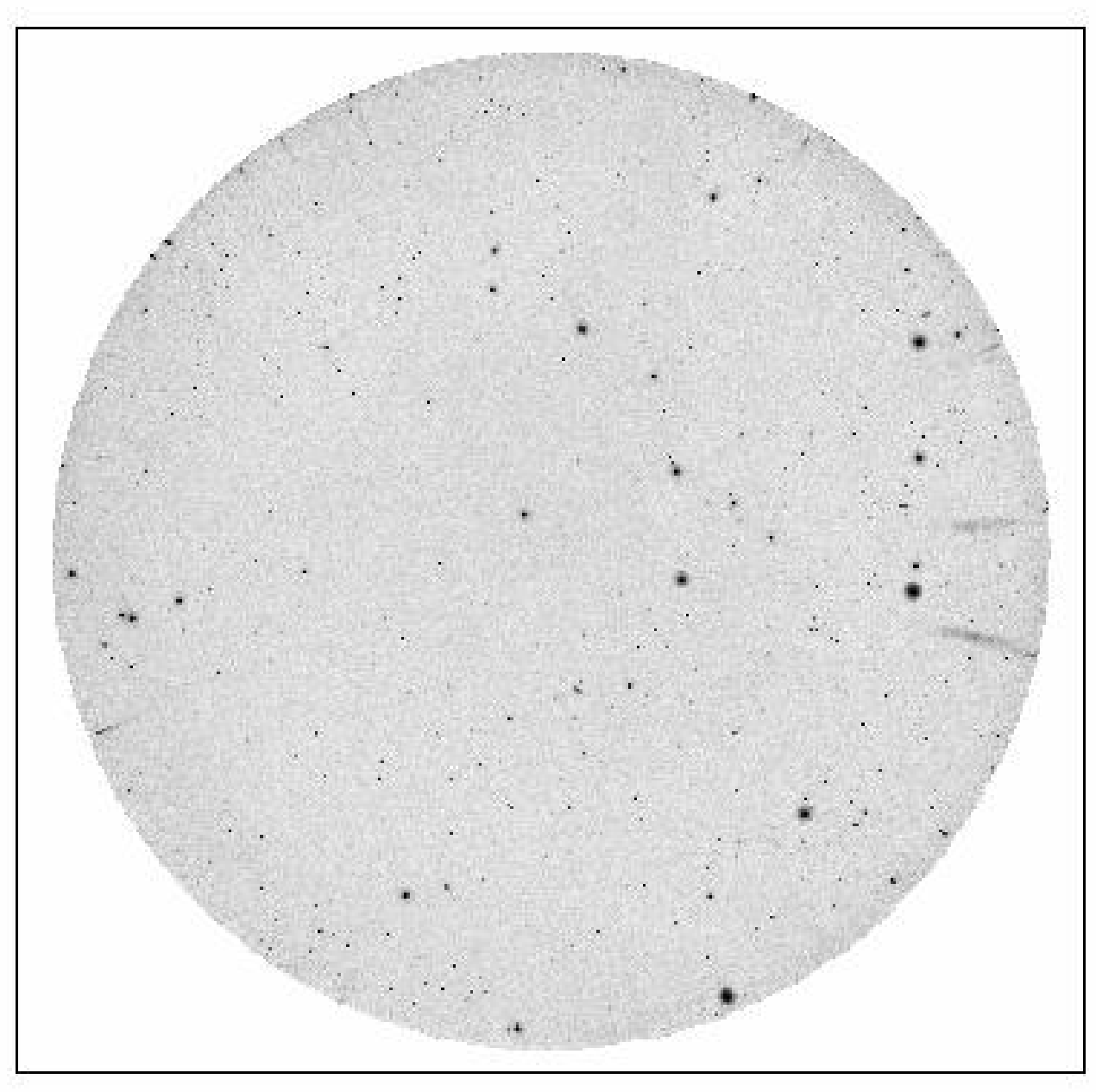}
\figcaption{Coadded images of FUV (top) and NUV (bottom) intensities observed by \galex. Note the very few and faint stars in both images.
\label{rawimage}}
\end{figure}

A single \galex\ observation is actually made up of a number of visits, each of approximately 30 minutes in length (limited by the duration of orbital night at this altitude) and possibly separated by several months. Although \galex\ is a photon counting instrument,  the standard \galex\ pipeline combines all observed events to produce a single image per channel per visit. We coadded the data from each visit to produce a single view of the observed region in each of the two channels (Fig. \ref{rawimage}).

These images contain both stars and the diffuse background but, because our focus is on the diffuse background, we have eliminated all the stars by cutting out a box around each source listed in the \galex\ merged catalog and then averaging the data into bins of 2\arcmin $\times$ 2\arcmin\ in size. Note that we eliminated the edge effects seen in all \galex\ observations before binning. These data represent the total diffuse radiation from all sources in this direction and in this observation. It is difficult for any imaging experiment to separate the different components of the diffuse radiation and, in the remainder of this section, we will discuss the possible sources of the background. It is important, however, to note that the direct stellar contribution (the sum of the counts from all the stars in the image) to the total signal in each channel is less than 7\%.

\subsection{Instrumental Effects}

Photon counting instruments such as the delay-line detectors used in \galex\ are intrinsically low noise and the instrumental dark count, due largely to cosmic rays and other fast particles, is only on the order of 20 and 60 counts per second over the FUV and NUV detectors, respectively \citep{Morissey2007}, corresponding to an effective diffuse background of about 5 \phunit. There are other instrument related effects which may give rise to artifacts in the background, most important of which is the flat field already applied to the data. Specifically to mitigate possible variations in the flat field, \galex\ observations are dithered over a 90\arcsec\ spiral leaving an empirical uncertainty of about 5\% in the flat field \citep{Morissey2007}. We would expect an even smaller variation in our data where we further bin over 2\arcmin\ square boxes.

\subsection{Airglow}

\begin{figure}
\epsscale{1.0}
\plotone{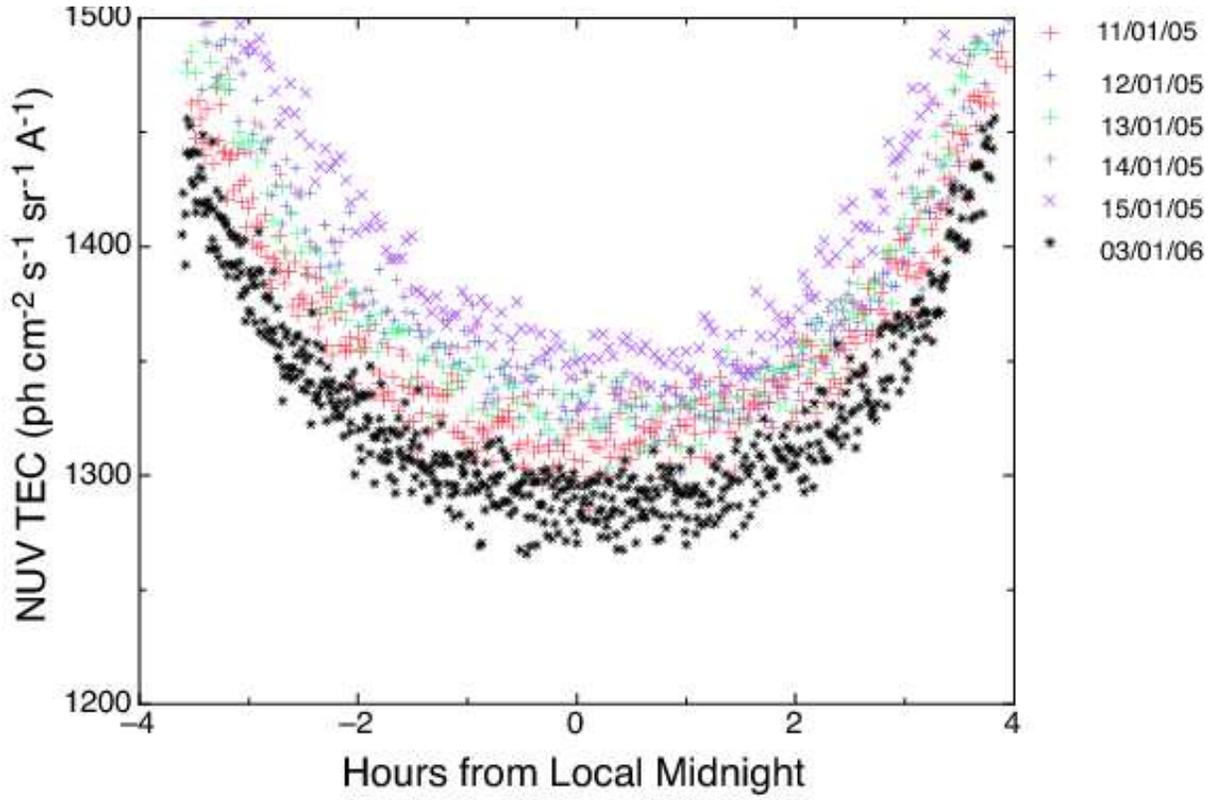}
\figcaption{Total count rate (TEC) converted into \phunit\ in the NUV channel as a function of local time for each of the visits. \label{NUVAirglow}}
\end{figure}

\begin{figure}
\epsscale{0.9}
\plotone{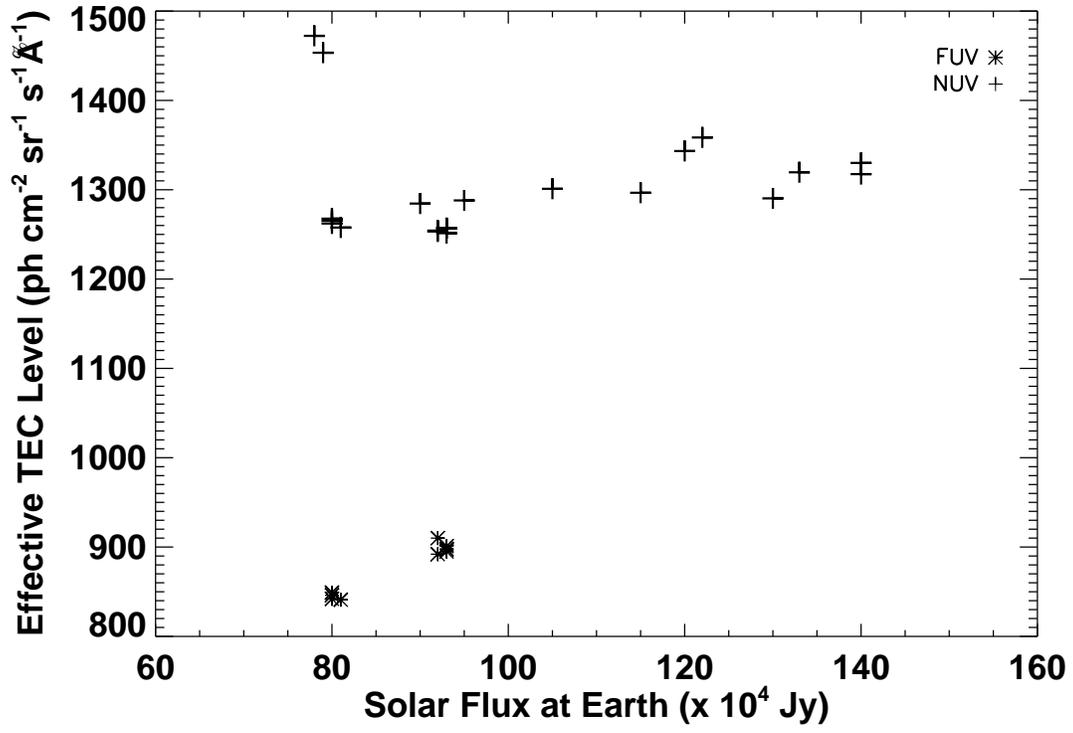}
\figcaption{Effective TEC level in each visit versus solar flux at the Earth. \label{SolarAirglow}}
\end{figure}

A significantly greater part of the emission is due to the resonantly scattered O\ {\rm \small I} lines at 1304, 1356, and 2471 \AA\ from the Earth's ionosphere with a contribution of between 100 and 200 \phunit\ in each of the two \galex\ bands \citep{Boffi2007}. Uniquely amongst the contributors to the diffuse radiation, the airglow will vary over the course of a single visit. Indeed, plotting the total count rate from the \galex\ TEC counter (Fig. \ref{NUVAirglow}) shows that the airglow in a given visit is a function of the local time, with a minimum at local midnight. However, there is also a variation between visits which appears to be roughly correlated with the level of solar activity\footnote{archived at http://www.dxlc.com by Jan Alvestad.} (Fig. \ref{SolarAirglow}) as would be expected from the resonantly scattered Solar photons in the Earth's ionosphere. There are two points in the NUV channel which blatantly violate this general trend. These are the only two in which the Sun angle is less than 90\degr\ and it is likely that the increased baseline is due to instrumental scattering of off-axis Solar light. These were not observed in the FUV because of the power issues at the time.

We have adopted an empirical method to estimate the airglow in each observation by setting the airglow contribution to 0 at local midnight on Jan. 3, 2006. The total airglow contribution to each visit is then the integral under the curve. These values are tabulated for the FUV channel in Table {\ref{FUVAirglow}} and imply an average contribution to the total diffuse background from the airglow of about 85 \phunit. A similar exercise for the  NUV channel yields an average contribution of 120 \phunit. Given the uncertainty in the baseline, we estimate that the total airglow contribution in our data is between 100 and 200 \phunit, consistent with the levels observed at the slightly lower  {\it Hubble Space Telescope} altitude of 600 km.

\subsection{Zodiacal light}

Moving out from the Earth, the next major contributor to the diffuse radiation field is the zodiacal light, sunlight scattered by interplanetary dust grains. The visible light distribution of the zodiacal light has been mapped as a function of helioecliptic coordinates by \citet{Leinert98} and we have estimated the contribution of the zodiacal light to the NUV band by assuming that the UV distribution is the same as that in the visible with a color close to 1; i.e., the ratio between the zodiacal light and the solar spectrum is the same at all wavelengths. With these assumptions,  the effective contribution of the zodiacal light is less than 10 \phunit\ in the FUV, where the solar spectrum vanishes, and 400 -- 470 \phunit\ in the NUV, depending on the date of the visit.

The main uncertainty in predicting the level of the zodiacal light is that the color may differ from unity. \citet{JM90} found a dependence of the color on the ecliptic latitude varying from 0.6 on the ecliptic plane to 1.2 at an ecliptic latitude of $60^{\circ}$, and 1 at the $50^{\circ}$ latitude of these observations. Taking this into account, we estimate a 10\% uncertainty in the zodiacal light or about 50 \phunit. It should be reemphasized that there is essentially no zodiacal light contribution to the FUV channel.

\subsection{\htwo\ Emission}

This \galex\ field is included in one of the regions observed by \citet{MHB90} with the Shuttle-based  {\it Berkeley Extreme Ultraviolet/Far Ultraviolet Shuttle Telescope} ({\it BEST}) --- their Target 6, a scan over Ursa Major. They detected emission from the Werner bands of molecular hydrogen over the entire observed region and speculated that it was due to an \htwo\ halo extending well outside the Sandage cloud. Extrapolating into the \galex\ bands with the models of \citet{McCandliss}, this emission implies a  uniform brightness of 135 \phunit\  over the FUV field with no contribution in the NUV channel. We will discuss the \htwo\ contribution to our observed emission below but note that we have found the level of the emission to vary between 0 and 250 \phunit\ over the field.

\subsection{Total Emission}

\begin{figure}
\epsscale{1.0}
\plotone{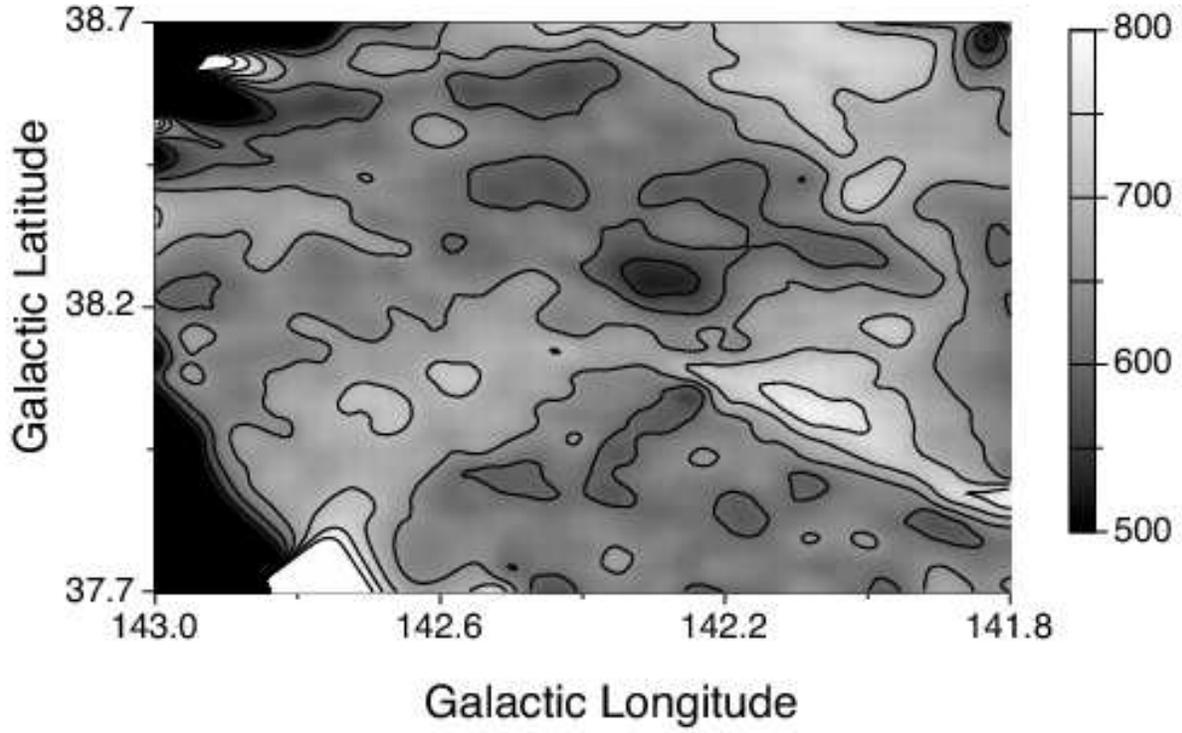}
\plotone{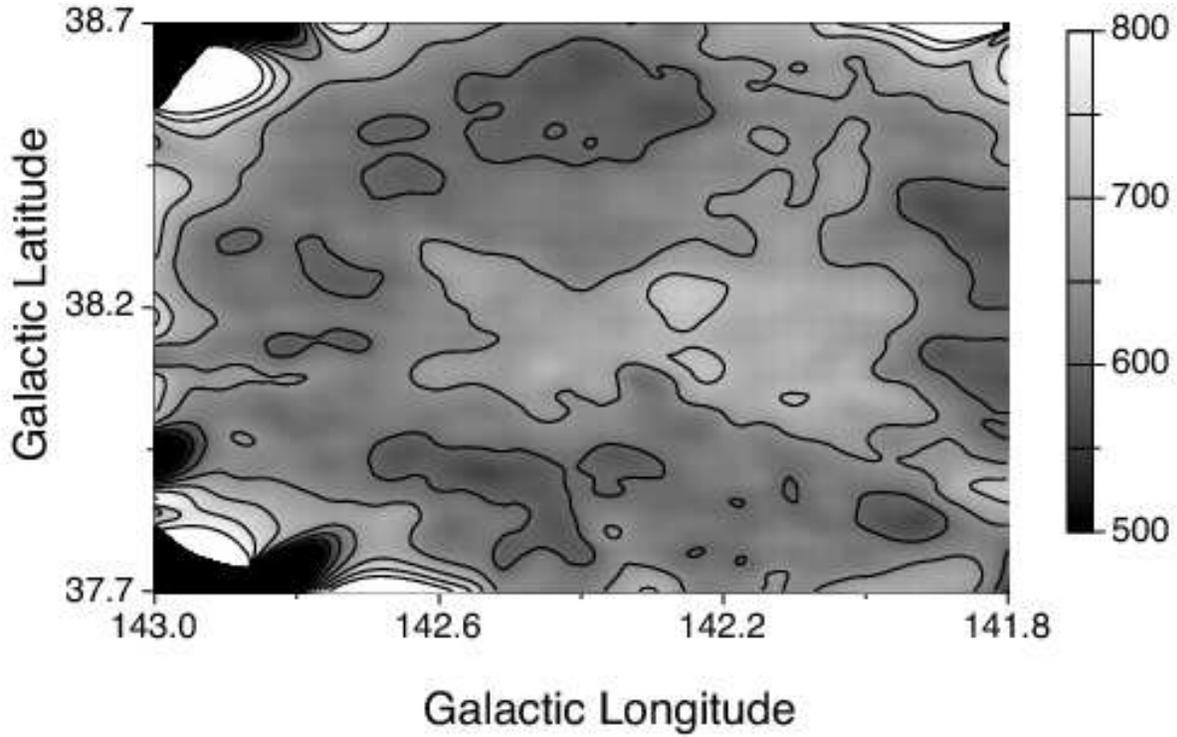}
\figcaption{\galex\ FUV (top) and NUV (bottom) diffuse images (with point sources removed) at a spatial resolution of 2\arcmin. The scale is in units of \phunit. \label{DiffuseImage}}
\end{figure}

The contribution of each of the sources of diffuse emission listed above to the total signal is tabulated in Table \ref{TotalEmission} implying a uniform contribution of about 90 \phunit\  in the FUV channel and 570 \phunit\ in the NUV, with an uncertainty of up to 100 \phunit\ in the baseline.  We have subtracted these uniform baselines from each channel and the resultant images are shown in Fig. \ref{DiffuseImage}. The remaining signal includes \htwo\ fluorescent emission, starlight scattered by the interstellar dust in the field and an extragalactic component, if any, and will vary across the image. We note that the foreground emission comprises about 15\% of the total signal in the FUV and close to 50\% of the total signal in the NUV, where the zodiacal light is of roughly the same magnitude as the Galactic sky background.

The uncertainty in the baseline is about 15\% of the total signal but it should be emphasized that this is not a pixel by pixel variation but rather a global uncertainty in the magnitude of the signal. The relative pixel-to-pixel uncertainty is largely determined by the uncertainty in the flat field and is less than 5\% \citep{Morissey2007}.

\section{RESULTS AND DISCUSSION}

\begin{figure}
\epsscale{0.9}
\plotone{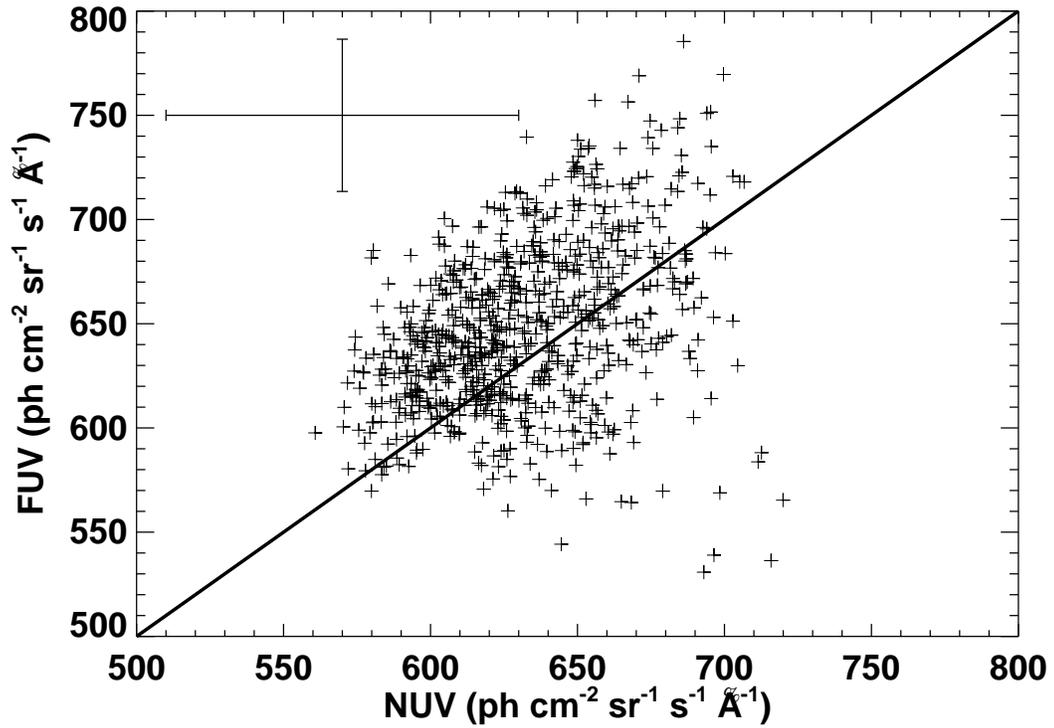}
\figcaption{FUV surface brightness with NUV surface brightness. Airglow has been subtracted from both channels with an additional zodiacal light subtraction from the NUV band. Representative error bars equal to about 5\% of the data values are shown. The linear correlation coefficient is 0.4 but, as noted in the text, we do not find a linear correlation. \label{fuv_nuv}}
\end{figure}

\begin{figure}
\epsscale{0.9}
\plotone{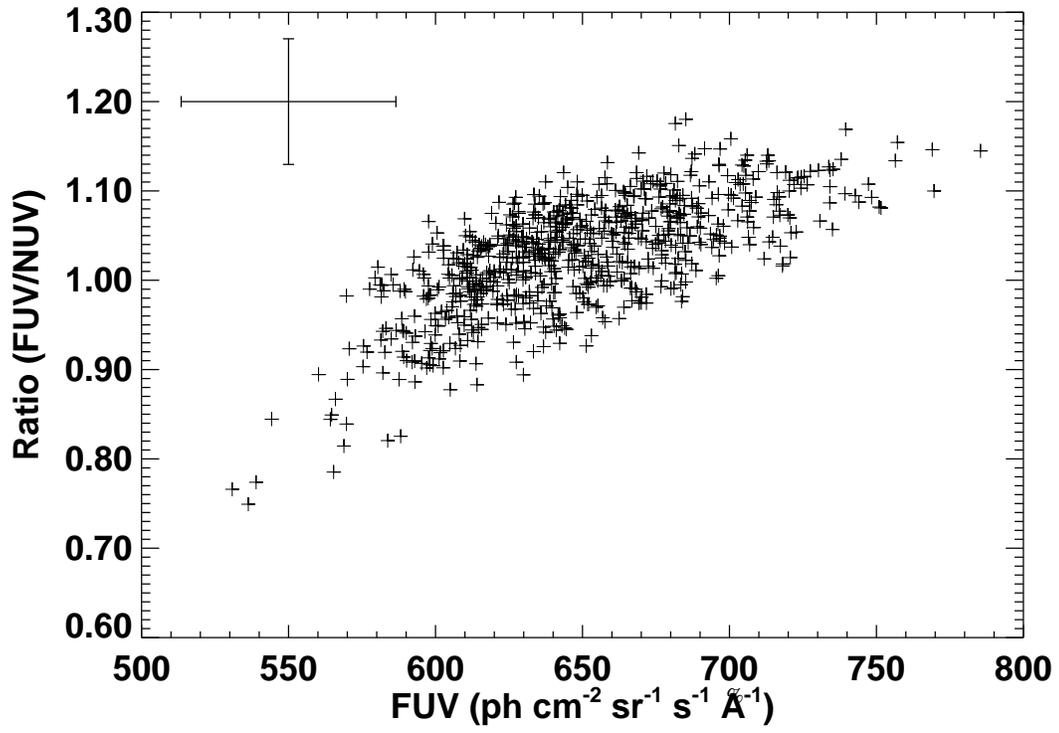}
\figcaption{Observed ratio between the FUV and NUV bands with representative error bars.The linear correlation coefficient is now 0.7.\label{fuv_nuv_ratio}}
\end{figure}

\begin{figure}
\epsscale{0.9}
\plotone{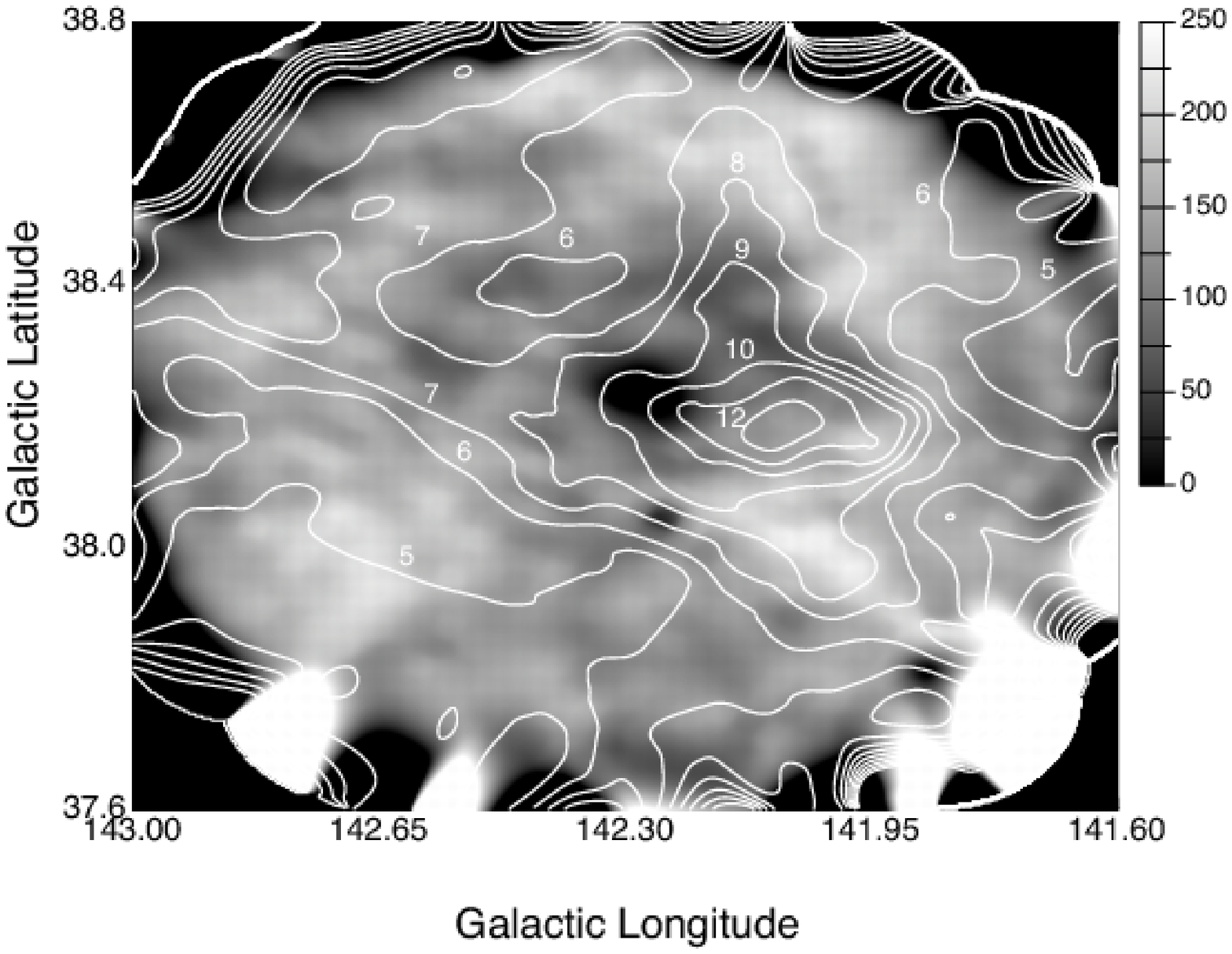}
\figcaption{The predicted level of \htwo\ emission in \phunit of the field. IRAS 100 \micron\ intensity contours in MJy sr$^{-1}$ are overplotted and marked.\label{h_two}}
\end{figure}

The emission in the FUV channel is strongly correlated with that in the NUV channel (Fig. \ref{fuv_nuv}), as would be expected from starlight scattered by interstellar dust. The illuminating ISRF is spectrally flat in this region and, purely coincidentally, the optical depth of normal Galactic dust is the same in both channels with the FUV rise in the extinction curve balanced by the 2175 \AA\ bump in the NUV. However, a closer glance at the figure shows that the ratio between the two bands (FUV/NUV) increases with the FUV emission (Fig. \ref{fuv_nuv_ratio}), suggesting a source which contributes to the FUV band but not the NUV. This is most probably due to molecular hydrogen fluorescence and we have estimated its magnitude and distribution by assuming that the ratio between the two bands to be fixed at 0.8 --- the minimum observed ratio --- and attributing the remainder to \htwo\ emission (Fig. \ref{h_two}). The absolute level may be uncertain by about 100 \phunit\, as discussed above, but this will not affect the morphology of the emission. 

\begin{figure}
\epsscale{0.9}
\plotone{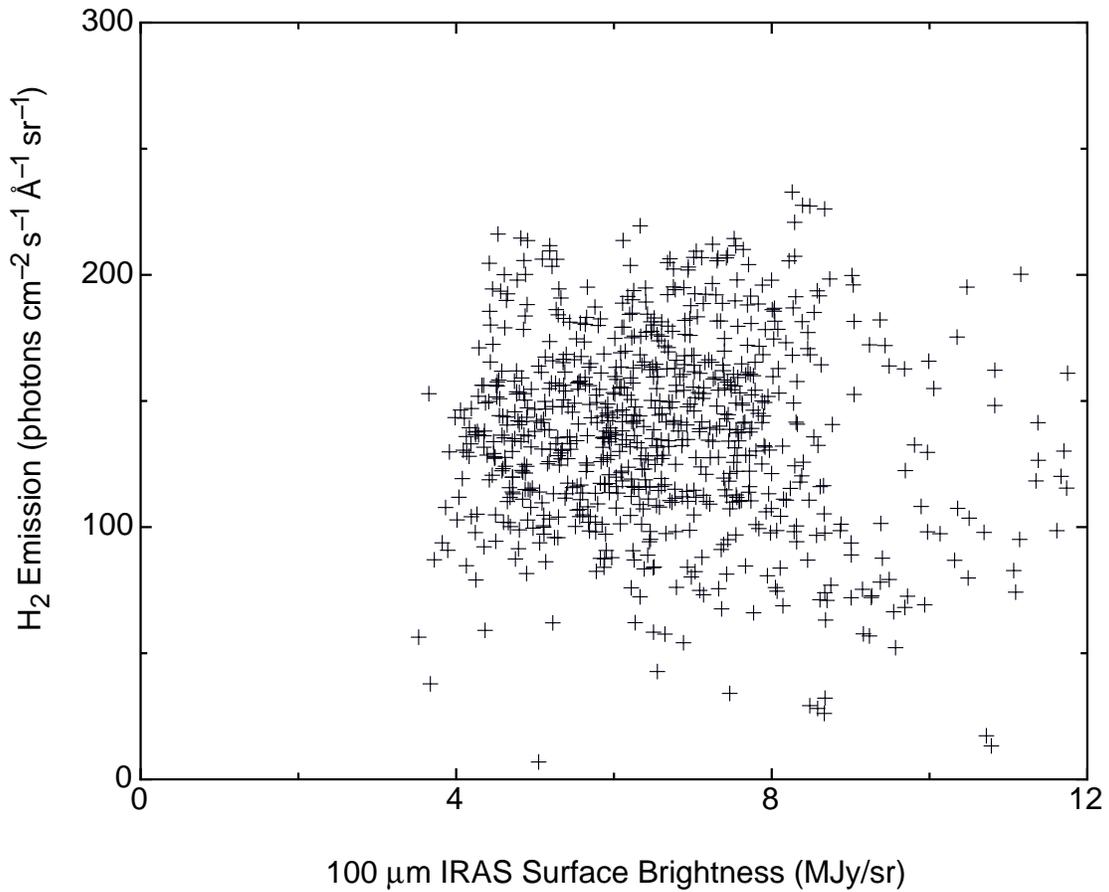}
\figcaption{The correlation between the IRAS 100 \micron\ intensity in MJy sr$^{-1}$ and the \htwo\ emission in \phunit\ is shown. With a correlation coefficient of -0.1, there is clearly no correlation between the two.\label{h_twoiras}}
\end{figure}

The level of the \htwo\ varies to a maximum of about 250 \phunit\ over the field with no correlation (Fig. \ref{h_twoiras}) with the \iras\ 100 \micron\ emission or, by extension, with CO  \citep{Weil}. These results are consistent with \citet{MHB90} who found from their {\it UVX/BEST} results that the \htwo\ emission extended much beyond the CO contours and suggested that it was due to an extended halo of molecular hydrogen. However, we do find a spatial variability in the emission to which they were not sensitive because of the speed of their scanning and the lower spatial resolution of the instrument. The remaining emission can be attributed to dust scattered radiation of 450 -- 650 \phunit\ in the FUV and 550 -- 750 \phunit\ in the NUV, with a spatial variation greater than that due to photon noise alone. Twenty years after {\it UVX}, it is difficult to correlate the \citet{MHB90} data with ours but it is clear that the two sets of observations have yielded consistent results.

\begin{figure}
\epsscale{0.9}
\plotone{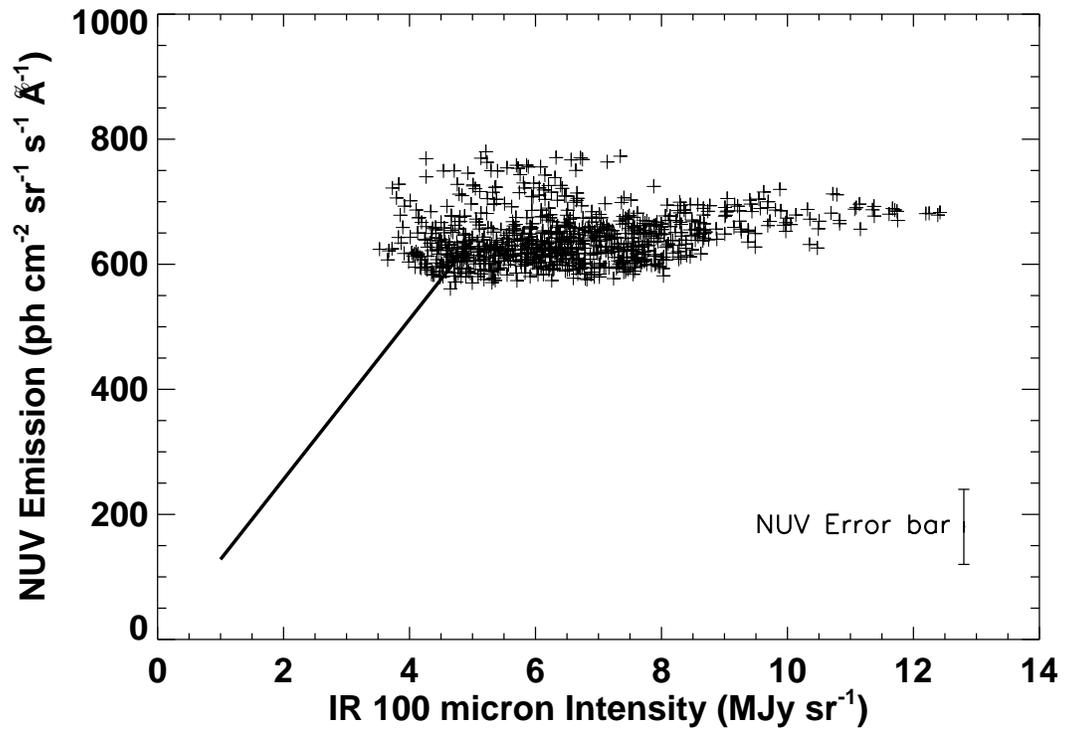}
\figcaption{IR-NUV correlation. The solid line is the observed slope of \citet{Haikala} from {\it FAUST} observations of the cloud G251.2+73 for the FUV (1400 - 1800\AA) region.\label{fuv_ir_ratio}}
\end{figure}

Our UV emission is flat relative to the 100 \micron\ emission (Fig. \ref{fuv_ir_ratio}) in contrast to the strong correlation found by \citet{Haikala} in {\it FAUST} observations of another cirrus cloud (G251.2+73.3). Both observations are of isolated clouds illuminated only by the ISRF but the optical depth is much greater in the Sandage region. Because the UV emission is due to scattering from the surface of the cloud while the IR is due to thermal emission from the entire volume of the cloud, because of the low optical depth in the infrared, we would expect a correlation between the UV and IR emission at low optical depths (in the UV) with the UV emission saturating as the 100 \micron\ emission increases. This is essentially what is seen in Fig. \ref{fuv_ir_ratio}, with the caveat that a more detailed modeling of both regions is required.

We have used our standard model for scattering from interstellar dust \citep{SNV05,SNV07} to derive the optical constants of the interstellar grains in this direction. This procedure uses the position and spectral type of the stars from the Hipparcos catalog to predict the radiation field at the location of the scattering dust, assumed to be at a distance of 120 pc. The stellar radiation is convolved with a Henyey-Greenstein \citep{Henyey41} scattering function for the grains to obtain the distribution and magnitude of the dust-scattered light in this direction. In our previous studies of the Ophiuchus region \citep{SNV05} and the Coalsack \citep{SNV07}, the scattering was from an optically thin region with a small number of illuminating stars. Here, many more stars contribute to the radiation field with an optically thick dust cloud and it was not practicable to run a Monte Carlo multiple scattering model for the entire region. Therefore we assumed single scattering throughout the region but applied a correction for multiple scattering based on limited Monte Carlo runs.

\begin{figure}
\epsscale{0.9}
\plotone{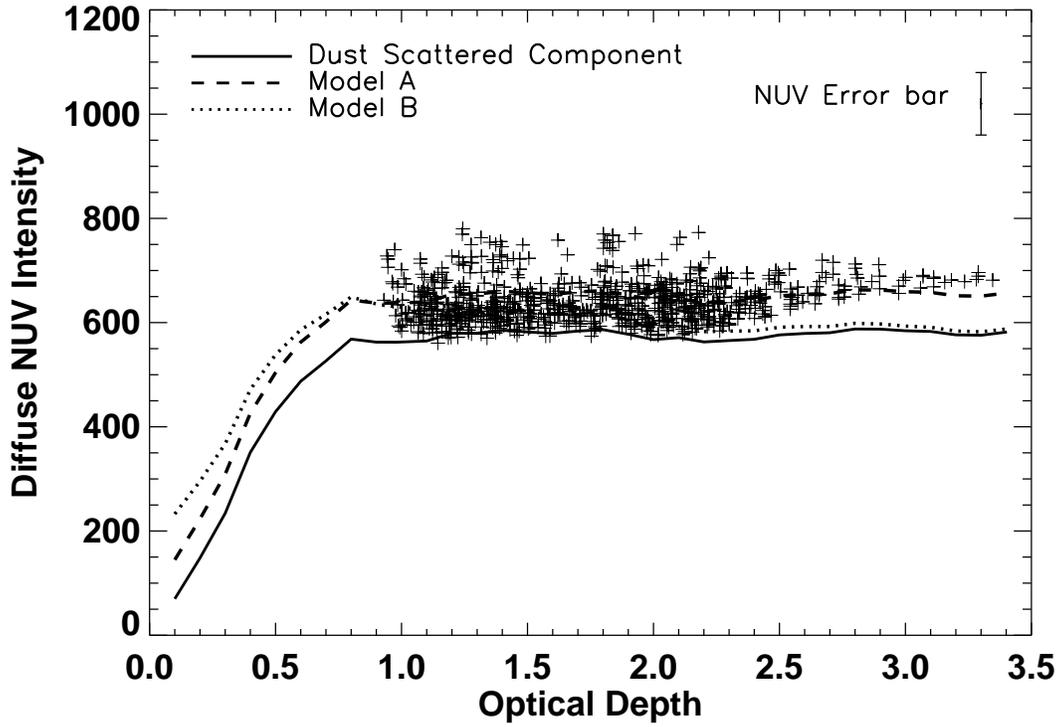}
\figcaption{Observed diffuse light with airglow and zodiacal light subtracted ('+') is plotted against the optical depth in the region. Overplotted lines are (a) model prediction of dust scattered star light (solid line), (b) dust scattered light plus a flat component representing an undersubtraction of airglow and zodiacal light (Model A - dashed line) (c) dust scattered light plus an extragalactic component (Model B - dotted line). A representative error bar for the NUV data is shown.\label{model}}
\end{figure}

Two representative model fits (A and B) to the NUV data with the airglow emission and zodiacal light subtracted are shown in Fig. \ref{model}. Both models include a dust scattered component corresponding to $a = 0.4; g = 0.7$ but model A (dashed line) includes a flat component representing the uncertainty in the baseline and model B includes an extragalactic component extincted by the foreground dust. The fit is much better in the case of model A with a reduced $\chi^{2}$ fit of 0.41  (with 862 degrees of freedom) as opposed to the 0.63 in the case of model B and we can formally rule out the presence of any extragalactic light at all on the basis of the $\chi^{2}$ values. However, our model only grossly represents the true situation as, for instance,  we have only approximated the effects of multiple scattering and have not considered the effects of clumping at all. These are the first observations of the diffuse background at such high spatial resolution and further work is required to fully understand its distribution.

The $\chi^{2}$ values are lower than one would expect given the preliminary nature of our model and are most likely indicative of overlarge error bars. As discussed above, the errors are dominated by the 5\% uncertainty in the flat field estimated through repeated observations of point sources at different positions on the detector \citep{Morissey2007} and may not be reflective of the uncertainty in the diffuse background at the larger spatial scales of our study. We are currently studying this with many more observations of other locations to deconvolve the instrumental effects from true sky variations.

\section{CONCLUSIONS}

We have obtained the highest spatial resolution images of the diffuse UV background to date with an effective spatial resolution of about 2\arcmin. These observations are in a region of moderate optical depth with $\tau\ > 1$ where  \citet{Sandage76} observed, and correctly identified, starlight scattered from interstellar dust. We have obtained \galex\ observations in both the FUV (1350 -- 1750 \AA) and NUV (1750 -- 2850 \AA) bands. After subtraction of the foreground airglow and zodiacal light, we were left with about 500 - 800 \phunit\ in both the FUV and NUV bands. The FUV/NUV ratio increased with increasing FUV emission suggesting the presence of \htwo\ fluorescent emission in the FUV ranging up to an integrated emission of about 250 \phunit. Our values are consistent with those of \citet{MHB90} in their scan of this region; however, we observe a spatial variability that was not possible with their observations.

When we link our data with the observations of G251.2+73.3 by \citet{Haikala}, we find that the scattered UV light increases linearly with the IR emission for low optical depths but saturates at optical depths near unity. This is as expected given that the thermal IR emission is from the entire volume of the cloud because of the low optical depth in the IR. We have used the same models as in our earlier studies of the diffuse background and have found that a dust scattered component with $a = 0.4; g = 0.7$ is consistent with the data; ie., the dust is strongly forward scattering with a moderate albedo. However, the data show an intrinsic scatter much greater than can be attributed to photon noise alone which must reflect structure at a spatial scale of at least 2\arcmin\, possibly due to variations in the ISRF and in the  distribution of the interstellar dust.

We are now extending our analysis to a much larger body of \galex\ observations, both our own and archival data. Such an investigation will help resolve some of the uncertainties in this work such as the contribution of airglow and the zodiacal light. Previous observations of the scattering by interstellar dust were on much larger spatial scales and indicated a local origin to much of the background; ie., both nearby hot stars and interstellar dust were required. \galex\ observations will allow us to probe the diffuse background at much higher spatial resolutions and thus to investigate the small scale structure of the ISM.

\acknowledgements
This research is based on data from the NASA's \galex\ GI program, GALEX GI1-005007-J092810p702308. \galex\ whch is operated for NASA by the California Institute of Technology under NASA contract NAS5-98034. We have also made use of NASA's Astrophysics Data System and the SIMBAD database operated at CDS, Strasbourg, France. We acknowledge the use of NASA's SkyView facility (http://skyview.gsfc.nasa.gov) located at NASA Goddard Space Flight Center. 

NVS is supported by a DST Young Scientist award. Support for RCH and LB was provided by NASA GALEX grants NNGO5GF19G and  NNGO6GF53G to the Johns Hopkins University.

We most gratefully thank Patrick Morrissey for extensive discussions and clarifications beyond the call of duty.

Facilities: \facility{\galex}.

\bibliography{ms}

\clearpage

\begin{deluxetable}{c c c}
\tablenum{1}
\tablecaption{OBSERVATION LOG}
\tablewidth{0pt}
\tablehead
{\colhead{} & \colhead{FUV} & \colhead{NUV}
}
\startdata
Wavelength range & 1350 - 1750 \AA & 1750 - 2850 \AA\\
$\lambda_{eff}$ &    1540 \AA & 2320 \AA\\
FOV 	& 1.27\degr 	& 1.24\degr \\
Image resolution &	4.2\arcsec & 5.3\arcsec \\
Exposure Time (s)&	14,821 &	35,210 \\
Number of Visits&	10 &	22 \\
RA     & \multicolumn{2}{c}{09 28 07}  \\
Dec    & \multicolumn{2}{c}{70 21 26}  \\
l      & \multicolumn{2}{c}{142.3} \\
b      & \multicolumn{2}{c}{38.2}
\enddata
\label{obslog}
\end{deluxetable}

\clearpage

\begin{deluxetable}{c c c c c c}
\tablenum{2}
\tablecaption{FUV AIRGLOW DETAILS}
\tablewidth{0pt}
\tablehead
{\colhead{Visit} & \colhead{FUV Minimum} & \colhead{AG$_{C}$\tablenotemark{2}} & \colhead{AG$_{V}$\tablenotemark{3}} & \colhead{Total AG} \\
\colhead{} & \colhead{} & \multicolumn{3}{c}{(\phunit)} 
}
\startdata
1  & 915 & 45 & 66 & 111 \\
2  & 921 & 45 & 70 & 115 \\
3  & 918 & 50 & 49 & 99  \\
4  & 917 & 50 & 44 & 94  \\
5  & 930 & 55 & 50 & 105 \\
6  & 912 & 55 & 65 & 120 \\
7  & 861 & 0  & 45 & 45  \\
8  & 869 & 5  & 51 & 56  \\
9  & 862 & 10 & 46 & 56  \\
10 & 866 & 5  & 32 & 37  \\
\hline
\multicolumn{4}{c}{Average Integrated Airglow} & 85
\enddata
\label{FUVAirglow}
\tablenotetext{2}{Relative to Visit 7 on Jan. 3, 2006.}
\tablenotetext{3}{Area under the curve above AG$_{C}$.}
\end{deluxetable}
\clearpage

\begin{deluxetable}{l c c}
\tablenum{3}
\tablecaption{CONTRIBUTORS TO DIFFUSE EMISSION}
\tablewidth{0pt}
\tablehead
{\colhead{Component} & \colhead{FUV} & \colhead{NUV}  \\
\colhead{} &  \multicolumn{2}{c}{(\phunit)} 
}
\startdata
Dark Count & 5 & 5\\
Airglow & 85 & 120 \\
Zodiacal Light & -- & 445 \\
\htwo Fluorescence & 0 - 250 & --\\
\hline
Total & 90 - 340 & 570
\enddata
\label{TotalEmission}
\end{deluxetable}

\clearpage
\clearpage

\clearpage

\clearpage

\clearpage

\clearpage

\clearpage

\clearpage
\clearpage

\clearpage

\clearpage

\end{document}